\documentclass[sigconf]{acmart}

\usepackage[colorinlistoftodos,prependcaption,textsize=tiny]{todonotes}

\usepackage{enumitem}
\usepackage{hyperref}
\usepackage{cleveref}

\usepackage{xspace} 
\newcommand{\header}[1]{{\noindent{\textbf{#1}}}}

\newcommand{\toolname}{AttentionFlow\xspace}
\newcommand{\Vevo}{\textsc{VevoMusic}\xspace}
\newcommand{\Wiki}{\textsc{WikiTraffic}\xspace}

\AtBeginDocument{%
  \providecommand\BibTeX{{%
    \normalfont B\kern-0.5em{\scshape i\kern-0.25em b}\kern-0.8em\TeX}}}

\copyrightyear{2021}
\acmYear{2021}
\setcopyright{rightsretained}
\acmConference[WSDM '21]{Proceedings of the Fourteenth ACM International Conference on Web Search and Data Mining}{March 8--12, 2021}{Virtual Event, Israel}
\acmBooktitle{Proceedings of the Fourteenth ACM International Conference on Web Search and Data Mining (WSDM '21), March 8--12, 2021, Virtual Event, Israel}
\acmDOI{10.1145/3437963.3441703}
\acmISBN{978-1-4503-8297-7/21/03}

\begin{document}
\fancyhead{}

\title{\toolname: Visualising Influence in Networks of Time Series \vspace{-4em}}

\author{ Minjeong Shin$^{1*}$, Alasdair Tran$^{1,2*}$, Siqi Wu$^{1*}$}
\author{Alexander Mathews$^{1}$, Rong Wang$^{1}$, Georgiana Lyall$^{1}$, Lexing Xie$^{1,2}$}

\affiliation{%
  \institution{$^1$ Australian National University $\qquad$ $^2$ Data61, CSIRO}
}
\email{{minjeong.shin,alasdair.tran,siqi.wu,alex.mathews,rong.wang,u6431454,lexing.xie}@anu.edu.au}

\renewcommand{\shortauthors}{Shin, et al.}



\settopmatter{printacmref=false, printfolios=false}


\begin{abstract}

\def\thefootnote{*}\footnotetext{These authors contributed equally to this
work.}\def\thefootnote{\arabic{footnote}}

The collective attention on online items such as web pages, search
terms, and videos reflects trends that are of social, cultural, and economic
interest. Moreover, attention trends of different items exhibit mutual
influence via mechanisms such as hyperlinks or recommendations. Many
visualisation tools exist for time series, network evolution, or network
influence; however, few systems connect all three. In this work, we present
\toolname, a new system to visualise networks of time series and the dynamic
influence they have on one another. Centred around an ego node, our system
simultaneously presents the time series on each node using two visual
encodings: a tree ring for an overview and a line chart for details. \toolname
supports interactions such as overlaying time series of influence, and
filtering neighbours by time or flux. We demonstrate \toolname using two
real-world datasets, \Vevo and \Wiki. We show that attention spikes in songs
can be explained by external events such as major awards, or changes in the
network such as the release of a new song. Separate case studies also
demonstrate how an artist's influence changes over their career, and that
correlated Wikipedia traffic is driven by cultural interests. More broadly,
\toolname can be generalised to visualise networks of time series on physical
infrastructures such as road networks, or natural phenomena such as weather and geological measurements.


\end{abstract}

\vspace{-1mm}

\vspace{-1mm}

\vspace{-1mm}


\begin{teaserfigure}
  \centering
  \includegraphics[width=0.98\linewidth]{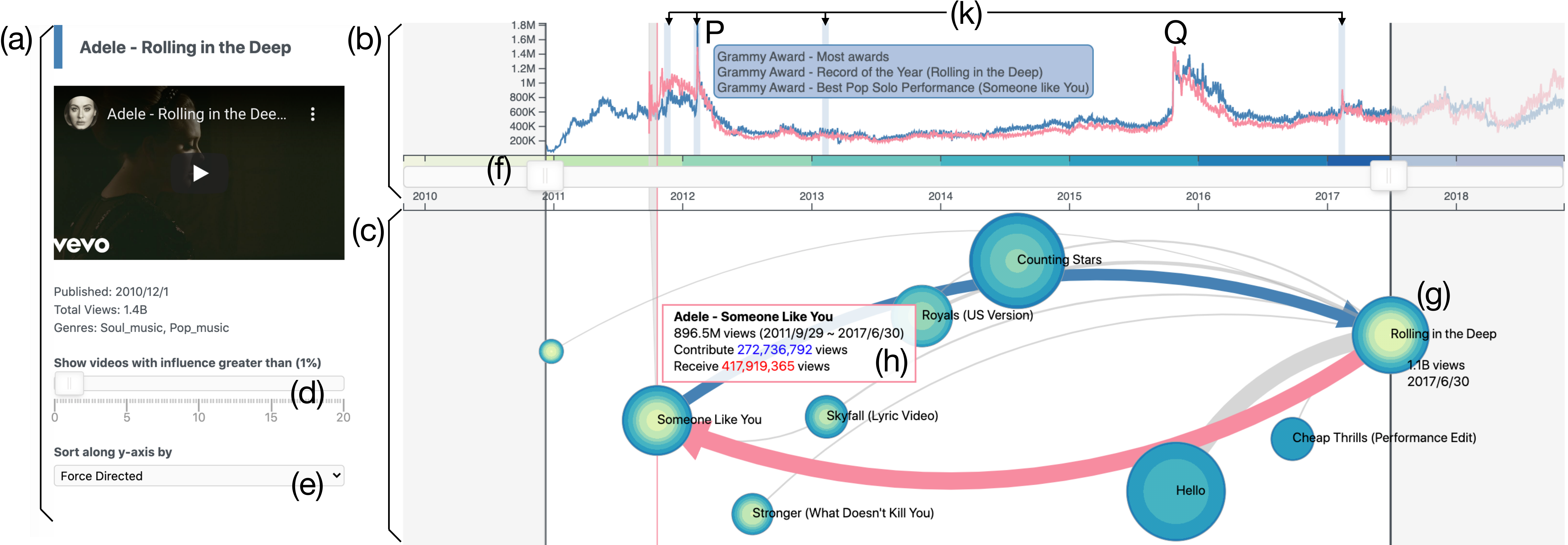}
  \vspace*{-2mm}
  \caption{\toolname visualises the attention series of an entity and the
  influence flowing over its ego network. The metadata view (a) provides
  high-level information about the ego node (g). The attention chart (b) presents
  two attention series of the ego and the hovered alter node, while the ego
  network (c) highlights the incoming and outgoing influence between them.
  Users can filter the alter nodes by setting the influence threshold (d), select a sorting criterion
  for the vertical axis (e), and define an observation window using the time
  slider (f).
  In this snapshot, we observe two spikes in the attention series of the
  music video {\em Rolling in the Deep} by Adele. The first spike (P) is
  related to the Grammy Awards of that year, while the second (Q) is due to the
  release of Adele's new song {\em Hello}. The remaining components are described in Section~\ref{sec:system} and~\ref{sec:vevo}.}
  \label{fig:teaser}
\end{teaserfigure}

\maketitle

\vspace{-1mm}
{\fontsize{8pt}{8pt} \selectfont
\textbf{ACM Reference Format:}\\
Minjeong Shin, Alasdair Tran, Siqi Wu, Alexander Mathews, Rong Wang, Georgiana Lyall, Lexing Xie. 2021. AttentionFlow: Visualising Influence in Networks of Time Series. In the Proceedings of the Fourteenth ACM International Conference on Web Search and Data Mining (WSDM '21), March 8–12, 2021, Virtual Event, Israel. ACM, New York, NY, USA, 4 pages. \url{https://doi.org/10.1145/3437963.3441703}}


\section{Introduction}
\label{sec:intro}

Attention series measure user interests towards an online item over time, such
as visits to a website, posts in a community, and searches on a topic. When
multiple attention series are related, they can exert mutual influence on one
another, forming an influence network. For example, traffic on a web page
influences linked pages~\cite{goodall2018situ}, posts in an online community
influences related communities~\cite{kumar2018community}, and the degree to
which a social group accepts an idea influences other social
groups~\cite{wang2016ideas}.
By visualising influence in networks of attention series, we can separate
exogenous and endogenous influences, identify attention series that have
substantial influence, and understand the network response to external shocks.

One line of related works aims to visualise snapshots of influence between
entities.
IdeaFlow~\cite{wang2016ideas} visualises how ideas flow within and across
multiple social groups by modelling the lead-lag relationships in text
clusters. Situ~\cite{goodall2018situ} combines flow visualisation and anomaly
detection to identify
suspicious network traffic. \citet{shin2019influence} propose a flower-like
metaphor for visualising the intellectual influence between academic entities.
These methods assume the network to be static and focus on visualising events
or flow rather than the complete time series. Another line of works focuses on
visualising network evolution. To visualise a scholar's temporal collaboration
graph, egoLines~\cite{zhao2016egolines} chooses a subway map metaphor, while
egoSlider~\cite{wu2015egoslider} uses a series of juxtaposed glyphs. Unlike our
work, the networks in these visualisations are presented as multiple snapshots
and nodes do not represent individual time series. Moreover, several approaches
have been developed for visualising multiple time series, including
stacking~\cite{havre2000themeriver} and
clustering~\cite{kwon2017clustervision}. \citet{bak2009spatiotemporal} propose
Growth Ring Maps where time series are presented as circles with radially
varying colours.
By contrast, we take on the challenge of simultaneously visualising multiple
time series, the degree of influence between these series, and the dynamic
network structure on which this influence acts. To the best of our knowledge,
no research has addressed these three angles at once.

We develop \toolname, a new system for
visualising the dynamic influence flow in an ego network with time-series
nodes. It is so named to reflect the common use case of visualising time series
of user attention in online networks.
\toolname visualises how the time series of a focal node (the ego) and its
direct neighbours (the alters) influence each other. For each node, we encode
the time series in two visual styles. A tree ring encoding allows many time
series to be compared simultaneously which is useful for identifying common
patterns and interesting dynamics. A line chart encoding allows a detailed
comparison of two time series. The chart also contains a set of markers, each
of which corresponds to a real-life event that could have influenced the time
series. The ego network is presented as a node-link diagram which shares the
time axis with the line chart, allowing network structure changes to be
visually compared with the time series. The position of an alter node on the
time axis indicates when it starts to influence the ego node. The two main
interactions are: (1) hovering over an alter node to show the detailed line
chart for that node, and (2) moving the time slider to select an observation
window. As the window shifts, the ego network structure changes, showing or
removing nodes depending on their influence flows and the dynamic graph
structure.

\toolname is preloaded with two 
large-scale networked time series datasets: a network of 31K music videos induced by the YouTube
recommender system~\cite{wu2019estimating} and a network of
366K Wikipedia web pages induced by hyperlinks~\cite{Tran2020Radflow}.
Our case studies demonstrate that \toolname helps explain sudden
surges in the attention series by visualising both external and internal
influences.

The main contributions of this work are:
\setlist{nolistsep}
\begin{itemize}[leftmargin=*, noitemsep]
	\item \toolname, a new interactive web app for visualising a dynamic network of
	mutually influencing time series\footnote{The demo is available at
\href{https://attentionflow.ml}{https://attentionflow.ml}.}.
	\item A novel combination of visual elements that simultaneously display
	the time series of nodes, the time-varying network structure, and the
	magnitude of influence flowing along edges.
	\item Case studies that use \toolname to interpret view count time series
	from YouTube and Wikipedia networks.
\end{itemize}


\section{Networks of attention series}
\label{sec:data}

\toolname is a general system for visualising multiple time series with a
network structure.
Each node in the network consists of a univariate time series
and other semantic metadata.
Directed edges model the flow of influence from source nodes to
target nodes, with a weights quantifying the strength of influence. The
network can be dynamic: edges can appear and disappear and
edge weights can vary over time.

We use \toolname to visualise two datasets: \Vevo~\cite{wu2019estimating}, a
YouTube recommendation network of music videos; and \Wiki
\cite{Tran2020Radflow}, a hyperlink network of Wikipedia articles. In \Vevo,
there are 31K nodes, each containing daily view counts for a music video
from its creation date to November 2018. If a video's recommendation list
contains the link of another video, we add an edge from the former to the
latter. This results in 45K edges. 
We create the artist-to-artist network by aggregating all videos from each of the 2,928 artists.
In \Wiki, we have 366K pages and 22M
edges with traffic data from July 2015 to June 2020. A
directed edge is added if a page appears as a hyperlink on another page.

Edge weights, which represents the influence strength of the source node on the
target node, are estimated by different methods in \Vevo and \Wiki . For \Vevo,
\citet{wu2019estimating} use ARNet, 
a regression model that learns edge-specific weights on the network over time. The displayed edge strength correspond to the estimated number of daily views flowing on the edge.  
For \Wiki,
\citet{Tran2020Radflow} use the multi-head attention scores from the neural
forecasting model RADflow as a proxy for influence. Both models empirically 
outperform previous state-of-the-art methods, thus giving credence to the
usefulness of these weights.


\section{Visualisation Design}
\label{sec:design}

\begin{figure*}[t]
    \centering
    \includegraphics[width=0.95\textwidth]{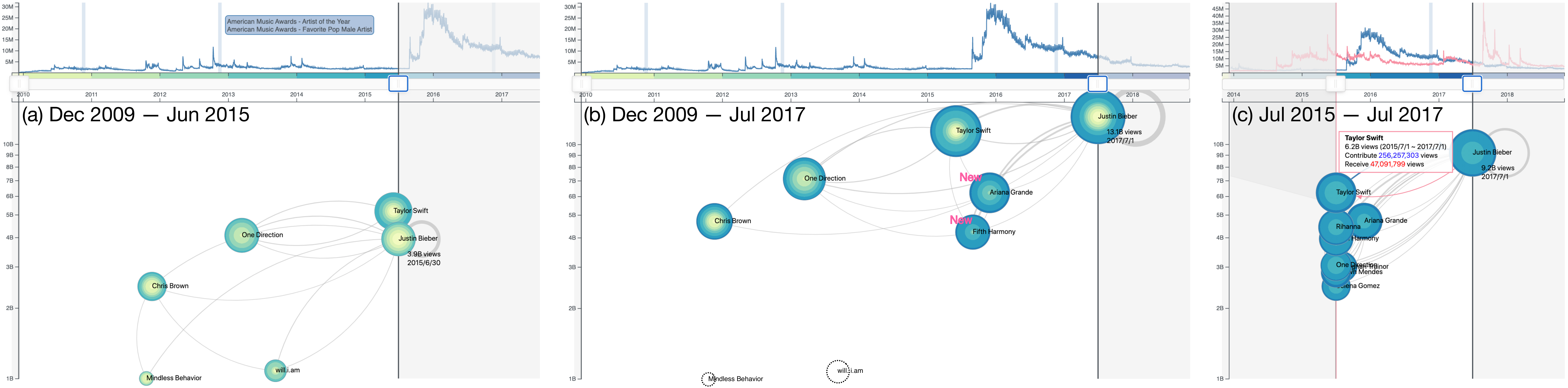}
    \vspace*{-2mm}
    \caption{Artist network page for Justin Bieber, showing neighbouring
    artists with more than 1\% influence. Nodes are sorted along the vertical axis by their total views. We show three network snapshots: (a)
    Dec 2009 -- Jun 2015, before
    Bieber published {\em Sorry}; (b) Dec 2009 -- Jul 2017, two years after it was published; and (c) Jul 2015 -- Jul 2017, non-overlapping two
    years after (a).}
    \label{fig:justinbieber_artistview}
    \vspace{-3mm}
\end{figure*}

\toolname focuses on the ego network rather than the entire network in order to
show how the temporal changes in attention correlate with the
influence around an ego node. We introduce two representations of nodes: an
attention chart and a tree ring.

The {\em attention chart} is a line chart for visualising the amount of
attention an ego attracts over time and providing detailed temporal patterns
such as peaks and valleys. It also allows users to compare trends of the ego
with an alter node by overlaying their attention time series on the same panel.
In \autoref{fig:teaser}b, the attention chart of an alter (pink) is overlaid on
top of the chart of the ego (blue), showing that {\em Someone Like You} has a
similar attention trend to {\em Rolling in the Deep}. We provide a time slider
to select an observation window, which plays an important role in determining
the sizes and colours of the elements described below.

The {\em tree ring} provides a coarse view of the attention series and allows
us to quickly compare the temporal trends of all nodes in an ego network. We
split the lifetime of a node into calendar years (though other time units could
also be used) each of which is assigned a colour. The size of the whole node
denotes the magnitude of attention
within the observation window. The size of each ring represents the amount of
attention within the corresponding year. In \autoref{fig:teaser}, {\em Rolling
in the Deep} has a timeline spanning over 8 years (2010--2017), hence its node
has 8 rings, with the innermost ring mapping to the attention in 2010 and
outermost ring mapping to 2017. Meanwhile, {\em Hello} has a shorter timeline
of 3 years. It has the biggest ring in the centre while the peripheral rings
are much thinner, indicating that it has an attention spike immediately after
its release and receives comparatively little attention in later years.

The {\em time-aligned ego network} emphasises the correlation between temporal
changes in the attention and influences from the alter nodes. It is drawn below
the attention chart so that they share the same timeline. We define the
{\em influencing time} of a node to be the time when the edge connecting it to
the ego has a normalised influence above a set threshold. We adapt the timeline
network layout introduced in EdgeMaps~\cite{dork2011edgemaps} to the setting
where the alter nodes are placed according to their influencing times. When the
influence threshold is set to zero, the influencing time becomes either the
start time of the observation window or the creation time of the node,
whichever comes later. Increasing the influence threshold will either move an
alter node to the right, or hide it if its influence falls below the threshold.
The user can interactively move the ego node along the timeline, defining the
right boundary of the observation window (\autoref{fig:justinbieber_artistview}). 
Edge width is calculated based on 
the amount of influence flowing from the source node to the target node,
normalised by the maximum influence in the ego network. 

%
\autoref{fig:justinbieber_artistview} (a--b) shows an example network in two
different time periods. 
In the earlier period (a), the ego node, {\em Justin
Bieber}, connects to five alter nodes because their edges have influence above
the threshold. After two years (b),
{\em Mindless Behavior} and {\em will.i.am} are no longer influential to the
ego and thus disappear. Meanwhile, two new alter nodes, {\em Ariana Grande} and
{\em Fifth Harmony}, appear in the network. Time-varying attributes such as
node attention (node size and vertical position) and the amount of influence 
(edge width) are visually changing from (a) to (b).


\section{Navigating \toolname}
\label{sec:system}

We implement \toolname to allow users to search for entities of interest, and
explore their temporal trends by interactively changing the time periods and
thresholds. The frontend is rendered in D3.js and the backend uses the Neo4j
graph database. Users can access the visualisation of an entity by searching
for its name (e.g. the video name or the artist name), or by clicking on the
corresponding node in the ego network of another entity.

\autoref{fig:teaser} presents the main visualisation layout for {\em Rolling in
the Deep} (g), a popular music video from Adele. It consists of three
components: a metadata view, an attention chart, and a time-aligned ego
network.
The {\em metadata view} (a) shows an ego's attributes such as the title, the
embedded snippet, the creation time, and the genres. Below the description of
the attributes, two controllers can be used to alter the network layout.
The influence slider (d) sets the influence threshold, defaulting to 1\%. The
drop-down box (e) provides five criteria for sorting nodes along the vertical
axis: force-directed (default), total views, incoming views, outgoing views, or
categories.
%
The {\em attention chart} (b) visualises the attention series of an ego (blue)
and a hovered alter node (pink).
The period between the alter's creation and influencing time is coloured in
grey, indicating the time that it takes to reach the chosen influence threshold
relative to the ego node. A time slider (f) is located on the horizontal axis
to select an observation window. The left handle changes the start time of the
observation, while the right handle changes the position of the ego (g). The
periods outside the selected time range are greyed out.
The {\em time-aligned ego network} (c) changes dynamically when users interact
with the time slider and the influence slider. When hovering over an alter, the
edges between the alter and the ego are highlighted, the alter's attention
series is revealed on the attention chart, and an information card (h) pops up.

\section{Explaining influence on \Vevo}
\label{sec:vevo}

\toolname can incorporate domain-specific datasets to help users interpret
trend changes. It draws the event indicators on the attention chart as shown in
\autoref{fig:teaser}(k). For example, \toolname uses a list curated from
Wikipedia to show music awards related to \Vevo. {\em Rolling in the Deep}
received a surge of attention at the beginning of 2012 (\autoref{fig:teaser}P).
Hovering on the event indicator near the peak tells us that the 54th Annual
Grammy Awards happened on the day before and Adele received three Grammys that
year. The alignment between the ego network and the attention chart helps us
see the correlated attention spikes easily. At the end of 2015
(\autoref{fig:teaser}Q), {\em Rolling in the Deep} received another sudden
surge after a long period of stable attention. Examining the network view, the
{\em Hello} node is aligned with the time of the spike. One can hover over {\em
Hello} to have the video's time series added to the attention chart. A thick
edge pointing from {\em Hello} to {\em Rolling in the Deep} is highlighted,
suggesting that the latter regained popularity in late 2015 because {\em Hello}
was released. Furthermore, Adele's other video, {\em Someone Like You},
exhibits a similar tree ring to the ego but with thick edges in both
directions. By hovering over it, we discover that {\em Someone Like You} has an
almost identical attention series to {\em Rolling in the Deep}, with both
benefiting from the release of {\em Hello}.


\toolname also provides the same visual layout for the artist network, which
helps us track both the attention an artist receives on YouTube and the
evolution of their influence network. \autoref{fig:justinbieber_artistview}
presents three time periods of {\em Justin Bieber} surrounding the release of his biggest hit song {\em Sorry} in 2015: (a) 2009-15, the six years before {\em Sorry}, (b) 2009-17 that includes two
years post {\em Sorry}, and (c) 2015-17, the two years after {\em Sorry}. We set the influence threshold to 1\% and sort the vertical axis
by total view counts
The publication of {\em Sorry} was a big turning point as he gained 9.2
billion views within two years of {\em Sorry}'s release. The visualisation reveals (perhaps surprisingly) that the three American
Music Awards in 2010, 2012, and 2016 did not have much affect on the attention
trends. In (a), five artists influenced Bieber. Two years after {\em Sorry} was
released (b), two artists disappeared
and are replaced by two more popular artists. The remaining alters and the ego
node moved upwards as they gained more views from (a) to (b). Bieber gained
more views than Taylor Swift in these two years. Comparing the non-overlapping
periods of (a) and (c) tells us that Bieber gained more attention in the
later two years than the previous six years. As he received more influence from
other artists in the later periods, Bieber's influence on his own videos
decreases as shown by a thinner self-loop.


\section{Explaining \Wiki} 
\label{sec:case}

\begin{figure}[t]
    \centering
    \includegraphics[width=0.47\textwidth]{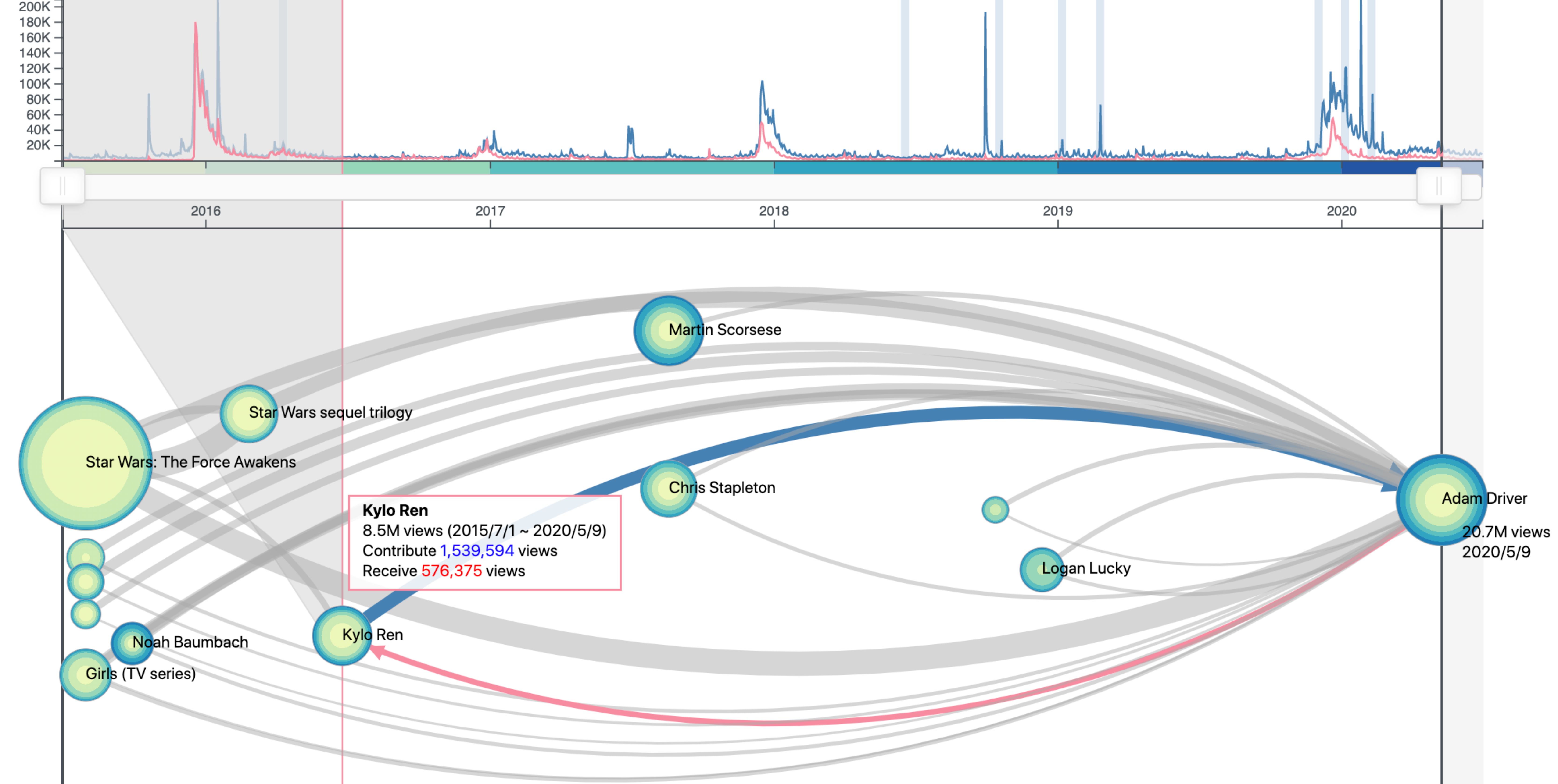}
    \vspace{-2mm}
    \caption{\toolname visualising the attention series of the Wikipedia page for
    {\em Adam Driver}, an actor who gained wider recognition for playing {\em Kylo Ren} in the {\em Star Wars} series. The influence threshold is set to 2\%.
    }
    \label{fig:wiki}
    \vspace{-3mm}
\end{figure}

\toolname can also visualise traffic flows between web pages.
\autoref{fig:wiki} presents the attention chart and the time-aligned ego
network of {\em Adam Driver}'s Wikipedia page, at an influence threshold of
2\%. We use film award data from IMDb to draw events related to the ego. {\em
Adam Driver} made his debut in 2010 but did not become widely recognised until
he was cast as {\em Kylo Ren} in {\em Star Wars: The Force Awakens} in 2015.
The attention series of {\em Adam Driver} (blue) is highly correlated with the
recent Star Wars films, getting a surge in traffic after the release of {\em
The Force Awakens} in 2015, {\em The Last Jedi} in 2017, and {\em The Rise of
Skywalker} in 2019. In particular, the thick edges between Driver and {\em The
Force Awakens} indicates substantial traffic flows between these two pages. The
page of {\em Kylo Ren} shows a similar attention series (pink) to {\em Adam
Driver}, especially before 2018. However, Driver received a higher number of
page views over time as he gained more popularity from films outside the Star
Wars franchise. Driver's attention chart also shows more peaks than {\em Kylo
Ren}, as these correspond not only to releases of Star Wars films but also to
the release of his other films and to his award nominations.

\section{Conclusion}
\label{sec:conclusion}

\toolname is a system for visualising networks of time series, with a novel
combination of multiple interactive visual elements including line charts, tree
rings, and dynamic ego networks.
We present three case studies to explore the influence networks
of videos and artists on YouTube, as well as Wikipedia traffic of cultural
entities. The system allows us to better understand the effects of new nodes,
the effects of external events, and to interpret the factors affecting an artist's career.
Future work includes deploying this system to networks of time
series beyond online attention, and improving network layout to reduce node overlap while
preserving the notion of time.


\header{Acknowledgements}
This research is supported in part by the Australian Research Council Project DP180101985.

\bibliographystyle{ACM-Reference-Format}
\bibliography{main}


 \newcommand{\noop}[1]{}
\begin{thebibliography}{12}


\ifx \showCODEN    \undefined \def \showCODEN     #1{\unskip}     \fi
\ifx \showDOI      \undefined \def \showDOI       #1{#1}\fi
\ifx \showISBNx    \undefined \def \showISBNx     #1{\unskip}     \fi
\ifx \showISBNxiii \undefined \def \showISBNxiii  #1{\unskip}     \fi
\ifx \showISSN     \undefined \def \showISSN      #1{\unskip}     \fi
\ifx \showLCCN     \undefined \def \showLCCN      #1{\unskip}     \fi
\ifx \shownote     \undefined \def \shownote      #1{#1}          \fi
\ifx \showarticletitle \undefined \def \showarticletitle #1{#1}   \fi
\ifx \showURL      \undefined \def \showURL       {\relax}        \fi
\providecommand\bibfield[2]{#2}
\providecommand\bibinfo[2]{#2}
\providecommand\natexlab[1]{#1}
\providecommand\showeprint[2][]{arXiv:#2}

\bibitem[\protect\citeauthoryear{Bak, Mansmann, Janetzko, and Keim}{Bak
  et~al\mbox{.}}{2009}]%
        {bak2009spatiotemporal}
\bibfield{author}{\bibinfo{person}{Peter Bak}, \bibinfo{person}{Florian
  Mansmann}, \bibinfo{person}{Halldor Janetzko}, {and} \bibinfo{person}{Daniel
  Keim}.} \bibinfo{year}{2009}\natexlab{}.
\newblock \showarticletitle{Spatiotemporal analysis of sensor logs using growth
  ring maps}.
\newblock \bibinfo{journal}{\emph{IEEE TVCG}} (\bibinfo{year}{2009}).
\newblock


\bibitem[\protect\citeauthoryear{D{\"o}rk, Carpendale, and Williamson}{D{\"o}rk
  et~al\mbox{.}}{2011}]%
        {dork2011edgemaps}
\bibfield{author}{\bibinfo{person}{Marian D{\"o}rk}, \bibinfo{person}{Sheelagh
  Carpendale}, {and} \bibinfo{person}{Carey Williamson}.}
  \bibinfo{year}{2011}\natexlab{}.
\newblock \showarticletitle{Edgemaps: Visualizing explicit and implicit
  relations}. In \bibinfo{booktitle}{\emph{VDA}}.
\newblock


\bibitem[\protect\citeauthoryear{Goodall, Ragan, Steed, Reed, Richardson,
  Huffer, Bridges, and Laska}{Goodall et~al\mbox{.}}{2018}]%
        {goodall2018situ}
\bibfield{author}{\bibinfo{person}{John~R Goodall}, \bibinfo{person}{Eric~D
  Ragan}, \bibinfo{person}{Chad~A Steed}, \bibinfo{person}{Joel~W Reed},
  \bibinfo{person}{G~David Richardson}, \bibinfo{person}{Kelly~MT Huffer},
  \bibinfo{person}{Robert~A Bridges}, {and} \bibinfo{person}{Jason~A Laska}.}
  \bibinfo{year}{2018}\natexlab{}.
\newblock \showarticletitle{Situ: Identifying and explaining suspicious
  behavior in networks}.
\newblock \bibinfo{journal}{\emph{IEEE TVCG}} (\bibinfo{year}{2018}).
\newblock


\bibitem[\protect\citeauthoryear{Havre, Hetzler, and Nowell}{Havre
  et~al\mbox{.}}{2000}]%
        {havre2000themeriver}
\bibfield{author}{\bibinfo{person}{Susan Havre}, \bibinfo{person}{Beth
  Hetzler}, {and} \bibinfo{person}{Lucy Nowell}.}
  \bibinfo{year}{2000}\natexlab{}.
\newblock \showarticletitle{ThemeRiver: Visualizing theme changes over time}.
  In \bibinfo{booktitle}{\emph{IEEE InfoVIS}}.
\newblock


\bibitem[\protect\citeauthoryear{Kumar, Hamilton, Leskovec, and Jurafsky}{Kumar
  et~al\mbox{.}}{2018}]%
        {kumar2018community}
\bibfield{author}{\bibinfo{person}{Srijan Kumar}, \bibinfo{person}{William~L
  Hamilton}, \bibinfo{person}{Jure Leskovec}, {and} \bibinfo{person}{Dan
  Jurafsky}.} \bibinfo{year}{2018}\natexlab{}.
\newblock \showarticletitle{Community interaction and conflict on the web}. In
  \bibinfo{booktitle}{\emph{WWW}}.
\newblock


\bibitem[\protect\citeauthoryear{Kwon, Eysenbach, Verma, Ng, De~Filippi,
  Stewart, and Perer}{Kwon et~al\mbox{.}}{2017}]%
        {kwon2017clustervision}
\bibfield{author}{\bibinfo{person}{Bum~Chul Kwon}, \bibinfo{person}{Ben
  Eysenbach}, \bibinfo{person}{Janu Verma}, \bibinfo{person}{Kenney Ng},
  \bibinfo{person}{Christopher De~Filippi}, \bibinfo{person}{Walter~F Stewart},
  {and} \bibinfo{person}{Adam Perer}.} \bibinfo{year}{2017}\natexlab{}.
\newblock \showarticletitle{Clustervision: Visual supervision of unsupervised
  clustering}.
\newblock \bibinfo{journal}{\emph{IEEE TVCG}} (\bibinfo{year}{2017}).
\newblock


\bibitem[\protect\citeauthoryear{Shin, Soen, Readshaw, Blackburn, Whitelaw, and
  Xie}{Shin et~al\mbox{.}}{2019}]%
        {shin2019influence}
\bibfield{author}{\bibinfo{person}{Minjeong Shin}, \bibinfo{person}{Alexander
  Soen}, \bibinfo{person}{Benjamin~T Readshaw}, \bibinfo{person}{Stephen~M
  Blackburn}, \bibinfo{person}{Mitchell Whitelaw}, {and}
  \bibinfo{person}{Lexing Xie}.} \bibinfo{year}{2019}\natexlab{}.
\newblock \showarticletitle{Influence flowers of academic entities}. In
  \bibinfo{booktitle}{\emph{IEEE VAST}}.
\newblock


\bibitem[\protect\citeauthoryear{Tran, Mathews, Ong, and Xie}{Tran
  et~al\mbox{.}}{2021}]%
        {Tran2020Radflow}
\bibfield{author}{\bibinfo{person}{Alasdair Tran}, \bibinfo{person}{Alexander
  Mathews}, \bibinfo{person}{Cheng~Soon Ong}, {and} \bibinfo{person}{Lexing
  Xie}.} \bibinfo{year}{2021}\natexlab{}.
\newblock \showarticletitle{Radflow: A recurrent, aggregated, and decomposable
  model for networks of time series}. In \bibinfo{booktitle}{\emph{WWW}}.
\newblock


\bibitem[\protect\citeauthoryear{Wang, Liu, Chen, Peng, Su, Yang, and Guo}{Wang
  et~al\mbox{.}}{2016}]%
        {wang2016ideas}
\bibfield{author}{\bibinfo{person}{Xiting Wang}, \bibinfo{person}{Shixia Liu},
  \bibinfo{person}{Yang Chen}, \bibinfo{person}{Tai-Quan Peng},
  \bibinfo{person}{Jing Su}, \bibinfo{person}{Jing Yang}, {and}
  \bibinfo{person}{Baining Guo}.} \bibinfo{year}{2016}\natexlab{}.
\newblock \showarticletitle{How ideas flow across multiple social groups}. In
  \bibinfo{booktitle}{\emph{IEEE VAST}}.
\newblock


\bibitem[\protect\citeauthoryear{Wu, Rizoiu, and Xie}{Wu et~al\mbox{.}}{2019}]%
        {wu2019estimating}
\bibfield{author}{\bibinfo{person}{Siqi Wu}, \bibinfo{person}{Marian-Andrei
  Rizoiu}, {and} \bibinfo{person}{Lexing Xie}.}
  \bibinfo{year}{2019}\natexlab{}.
\newblock \showarticletitle{Estimating attention flow in online video
  networks}.
\newblock \bibinfo{journal}{\emph{ACM CSCW}} (\bibinfo{year}{2019}).
\newblock


\bibitem[\protect\citeauthoryear{Wu, Pitipornvivat, Zhao, Yang, Huang, and
  Qu}{Wu et~al\mbox{.}}{2015}]%
        {wu2015egoslider}
\bibfield{author}{\bibinfo{person}{Yanhong Wu}, \bibinfo{person}{Naveen
  Pitipornvivat}, \bibinfo{person}{Jian Zhao}, \bibinfo{person}{Sixiao Yang},
  \bibinfo{person}{Guowei Huang}, {and} \bibinfo{person}{Huamin Qu}.}
  \bibinfo{year}{2015}\natexlab{}.
\newblock \showarticletitle{egoSlider: Visual analysis of egocentric network
  evolution}.
\newblock \bibinfo{journal}{\emph{IEEE TVCG}} (\bibinfo{year}{2015}).
\newblock


\bibitem[\protect\citeauthoryear{Zhao, Glueck, Chevalier, Wu, and Khan}{Zhao
  et~al\mbox{.}}{2016}]%
        {zhao2016egolines}
\bibfield{author}{\bibinfo{person}{Jian Zhao}, \bibinfo{person}{Michael
  Glueck}, \bibinfo{person}{Fanny Chevalier}, \bibinfo{person}{Yanhong Wu},
  {and} \bibinfo{person}{Azam Khan}.} \bibinfo{year}{2016}\natexlab{}.
\newblock \showarticletitle{Egocentric analysis of dynamic networks with
  egolines}. In \bibinfo{booktitle}{\emph{ACM CHI}}.
\newblock


\end{thebibliography}

\end{document}